\documentstyle[aps,prl,multicol,epsfig]{revtex}
\begin{document}
\title{Structure Formation, Melting, and the 
Optical Properties of Gold/DNA Nanocomposites: Effects of Relaxation Time}
\author{Sung Yong Park and D. Stroud}
\address{
Department of Physics,
The Ohio State University, Columbus, Ohio 43210}
\date{\today}
\maketitle
\begin{abstract}

We present a model for structure formation, melting, and optical
properties of gold/DNA nanocomposites.   These composites consist of
a collection of gold nanoparticles (of radius 50 nm or less) which
are bound together by links made up of DNA strands.  In our structural
model, the nanocomposite forms from a series of Monte Carlo steps,
each involving reaction-limited cluster-cluster aggregation (RLCA) followed by
dehybridization of the DNA links.   These links form with a probability 
$p_{eff}$ which depends on temperature and particle radius $a$.  The final
structure depends on the number of monomers (i.\ e. gold nanoparticles) $N_m$,
$T$, and the relaxation time.  At low temperature,
the model results in an RLCA cluster.  But after a long enough relaxation
time, the nanocomposite reduces to a compact, non-fractal cluster.
We calculate the optical properties of the resulting aggregates using
the Discrete Dipole Approximation.   Despite the restructuring, the
melting transition (as seen in the extinction coefficient at wavelength 
$520$ nm) remains sharp, and the melting temperature $T_M$ increases
with increasing $a$ as found in our previous percolation model.
However, restructuring increases the corresponding
link fraction at melting to a value well above the percolation threshold.
Our calculated extinction cross section agrees qualitatively with experiments on
gold/DNA composites.  It also shows a characteristic
``rebound effect,'' resulting from incomplete relaxation, which has also
been seen in some experiments.  We discuss briefly how our results relate
to a possible sol-gel transition in these aggregates.

\end{abstract}


\begin{multicols}{2}

\section{Introduction}

The optical properties of metallic nanoparticles have been 
investigated intensively for many years~\cite{bohren,kreibig}.
Recently, this work has extended to so-called functional 
metallic nanoparticles, which may have a variety of
novel and useful optical and mechanical properties~\cite{sanchez}.
Among these, there has been particular interest in the
DNA-modified gold nanoparticle (gold/DNA nanocomposite) system.  
This is a material consisting of gold nanoparticles to which specific kinds of
organic molecules (e.\ g., noncomplementary oligonucleotides capped 
with thiol groups) can be attached.  These materials can be
produced in a variety of structures using a strategy of 
nanoparticle self-assembly\cite{mirkin,alivisatos,winfree,zanchet,nelson}.  
They may also be useful
for selective biological detection, by making use of the
optical and electrical sensitivity of their 
aggregates~\cite{elghanian,bruchez,chan,taton,sjpark,cao}.  

Numerical model calculations of the optical properties of DNA 
modified gold nanoparticle aggregates show general agreement with 
experiments.  For example, both experiments and calculations show
that (i) for isolated gold nanoparticles in suspension, there is a 
strong surface plasmon absorption in the visible, arising from
oscillations of electronic charge in the gold nanoparticles; 
and (ii) this absorption maximum broadens and red-shifts when the 
cluster radius becomes comparable to the 
wavelength~\cite{lazar,lazar1,storhoff1}.   

Although the DNA molecules absorb primarily in the ultraviolet,
they nonetheless play a central role in the optical properties
of these aggregates in the visible, because they strongly
influence the {\em structure} of the gold/DNA aggregates.  For
example, the DNA tends to form multiple links between individual
gold nanoparticles.  These multiple links appear to account
for some key structural features of the 
aggregates\cite{park1,park2,others}.
In particular, the melting transition for a gold
nanoparticle aggregate has a much narrower temperature
width, and occurs at a substantially higher temperature T, than
that of a single DNA duplex ~\cite{elghanian,drukker}.    
In addition, the presence of multiple links leads to a
natural explanation for the dependence of the aggregate
melting temperature, denoted $T_M$ on particle size~\cite{others,kiang}.

In the present work, we extend our previous model 
calculations\cite{park1,park2} to take
into account the {\em dynamics} of aggregate formation, and how
these dynamics affect the aggregate optical properties.   In
our previous work, the melting of the aggregates was treated using
a purely statistical criterion.  Specifically, the aggregates were
formed by removing DNA links between gold nanoparticles 
on a simple cubic lattice, with a suitable, temperature-dependent
probability.  In the limit of a large aggregate, the melting
transition occurs, in this model, when the fraction of links
falls below the percolation threshold $p_c$ for the lattice 
considered.  The calculated  aggregate optical properties 
are found to change dramatically when this threshold is
passed, in good agreement with experiment.

To improve on this approach, we describe below a model for
structure formation which starts from isolated gold nanoparticles.
Our model takes into account two important features of the
structure formation:  the diffusion of 
nanoparticles, or clusters of nanoparticles, through the solvent to 
form a cluster, and the chemical reaction between DNA chains
which produces the links between the nanoparticles.  

Our structural model leads to a wide range of possible aggregate
morphologies, depending on the temperature.
Corresponding to these morphologies is a broad range of possible
optical properties.  In this paper, we will present numerical
results for both the structural and optical properties of these
nanocomposites over a typical range of parameters.  
For physically reasonable parameters, our numerical results
are in good agreement with experiment.   We will present a
qualitative interpretation of these results, and compare them
to our earlier, purely percolation model.

The remainder of this paper is organized as follows.
In Section II, we describe our structural model for gold/DNA
nanocomposites, and explain how it is implemented numerically.
In Section III, we review the Discrete Dipole Approximation,
which is the method used to calculate the optical properties of this system;
we also discuss various technical details needed to treat the
irregular clusters which emerge from the structural model of
Section II.  In Section IV, we present our numerical results for
both the structural and the optical properties of the aggregates.
Finally, in Section V, we summarize our results, interpret them
in terms of the expected behavior of typical gold/DNA nanocomposites,
compare our results to available experiments, 
and discuss their possible implications for future work.

\section{Structural Model and its Numerical Implementation}

In this Section, we describe our structural model for the formation
of gold/DNA aggregates.   
We start by describing the expected aggregation behavior 
at low T.  Following this, we present our full structural model and 
its numerical implementation for arbitrary T.

\subsection{Aggregation at Low $T$.}

Before discussing the aggregation of gold/DNA nanocomposites
at low T, we first consider the aggregation behavior of
other typical colloids, i.\ e. suspensions of small solid particles in a liquid 
solvent.  As the individual colloidal particles stick together to 
become clusters, these clusters themselves diffuse through the 
solvent, and continue to collide and aggregate.
This behavior is an example of cluster-cluster aggregation.  

If bonding between two colloidal particles is irreversible,
the final aggregate generally results from one of two processes: 
reaction-limited aggregation (RLA) or diffusion-limited 
aggregation(DLA). 
If there is a repulsive energy barrier between two 
approaching colloidal particles, the resulting process is expected to be a 
reaction-limited aggregation, because the reaction barrier is the limiting 
step in cluster growth~\cite{lin}.  
By contrast, if there were no barrier, the aggregation would be 
dominated by diffusion processes, and the resulting clusters 
should exhibit the features of DLA.
DLA clusters, like RLA clusters, are fractals, but they have a 
substantially lower $d_f$ than the RLA clusters.  
For forming both types of fractal clusters, however, 
the irreversibility of binding is important.  Without 
irreversibility, the final cluster is likely to become compact and 
non-fractal~\cite{reversible,terao}. 

We now provide a possible justification for considering the growth of the 
colloidal cluster in a dilute solution of DNA-modified gold particles 
at very low T as reaction-limited cluster-cluster aggregation (RLCA).
Here, we should note that for the DNA-modified gold nanoparticle 
system, the mechanism for binding two DNA-modified gold nanoparticles differs 
from the ordinary colloid aggregation process discussed above, 
because binding can occur only if the DNA hybridizes, i.\ e., if the
two DNA single strands on different gold particles and one linker 
single DNA strand undergo a reversible chemical reaction to form a double 
strand.  However, for two reasons, RLCA may still be a plausible growth 
mechanism for the gold cluster aggregation in a dilute solution at very low T. 
First, at the very low T, once the DNA hybridization occurs and
the two gold particles do stick together, they rarely unbind, since 
thermal fluctuation cannot provide enough energy to break them apart. 
Thus, the aggregation precess is {\em irreversible}.
Secondly, in case of DNA hybridization, this reaction barrier
can actually be experimentally observed~\cite{bloomfield}.
Because of this barrier, nanoparticles must collide numerous
times before two particles can stick together, 
since DNA hybridization takes a finite time.  

As T increases, the above argument, based on the irreversible binding, 
is no longer valid, since DNA dehybridization may easily take place.
Thus, the restructuring of clusters is crucial for understanding 
the cluster morphology and the melting transition. 
We now describe a structural model which includes the effects of this 
restructuring on the cluster morphology.

\subsection{Model for Structure Formation at General Temperature}

\subsubsection{Description of the Model}

At finite T, cluster restructuring is sensitive to the relative 
magnitudes of two times scales, which we denote 
$\tau_{bind}$ and $\tau_{dehyb}$.  $\tau_{bind}$ 
is the time elapsed when two clusters meet and attempt 
to form a link, while $\tau_{dehyb}$ is the time
needed for a a double DNA strand to dehybridize
into two single DNA strands under the influence of
thermal fluctuations.  At low T,  
$\tau_{dehyb} \gg \tau_{bind}$~\cite{bloomfield}.  Hence,
there will be many cluster-cluster binding events in the
time required for a one dehybridization. 
In the present model, 
we will assume that $\tau_{dehyb} \gg \tau_{bind}$ at {\em  all} T.  
With this assumption, we can consider aggregation and 
dehybridization processes separately.   
That is, we can establish a model in which, first
the monomers undergo aggregation and form one single large
cluster, and, next, the cluster breaks into smaller parts
via a suitable cluster unbinding process due to DNA 
dehybridization~\cite{note}.

Our aggregation procedure is carried out by a model of 
RLCA~\cite{rlca}, as justified in the previous subsection.  
In this algorithm,
the system is assumed to be made up initially of a large
number $N_m$ of ``monomers'' (each consisting of a single gold
nanosphere of radius $a$).  In the first step of the aggregation, 
we choose two monomers at random, then place them at two different 
points chosen randomly on the sites of a simple cubic lattice of 
lattice constant $\ell$ and edge $L \equiv N\ell$ in $d$ dimensions
($d$ = 3 in all of our simulations), with free boundary conditions.  
If the particles happen to be placed on adjacent sites, they are 
assumed to
form a two-particle cluster, and are removed from the lattice.
If they are placed on non-adjacent sites, the procedure is 
repeated until they 
do form a cluster.  In the next step, two of the $N_m - 1$ 
clusters, chosen at 
random, are placed in random, but non-overlapping positions and 
random orientations on the lattice.  If they are adjacent, 
the two clusters are assumed to merge
and form a larger cluster; otherwise, the procedure is repeated 
until the number 
of clusters is again reduced by one.  In the n$^{th}$ step, the 
same procedure is carried with two clusters randomly chosen from the now 
$N_m - n + 1$ clusters.  Eventually, this procedure 
leads to the formation of a single large cluster.

Once our aggregation procedure is finished, the next step in the procedure
is to simulate cluster unbinding due to DNA dehybridization. 
Here we use a percolation algorithm, where the key parameter is $p_{eff}$, 
the probability that a given link between two monomers in the cluster remains 
the same without breaking. Note that this $p_{eff}$ is obviously related to 
DNA hybridization, and thus depends on temperature. (The exact relation of 
$p_{eff}$ to DNA hybridization is discussed in the next subsection.)   
With probability $1 - p_{eff}$, we randomly remove links in the cluster, which
has formed from the aggregation procedure.
After random removal of links, the cluster may separate several clusters,
and thus we use a simple computer algorithm in order to identify the separate 
clusters~\cite{stauffer}.
If the resulting aggregate consists of two or more clusters, 
these aggregates are placed in random, but non-overlapping, 
positions and orientations on the large cubic lattice. 
This configuration will be used for the calculation of optical properties.

The two procedures described above are defined as one single 
Monte Carlo  (MC) ``step.'' To generate the final configuration, we 
carry out a series of $N_{MC}$ MC steps, each involving aggregation and cluster 
unbinding.   The end result is one or more final clusters.

The structure and the number of the final cluster or 
clusters depend on $p_{eff}$.  
If $p_{eff} = 1$, the final result is simply a single cluster 
formed from the aggregation part of the first MC step, which 
for large enough $N_m$, is known to be a RLCA fractal, 
with fractal dimension $d_f = 2.1$ if $d$ = 3~\cite{rlca}.  
For $p_{eff}<1,$ two types of transitions can take place in this system:
a ``sol-gel transition'' and a ``melting transition.''
If $p_{eff}$ is slightly less than  1, the restructuring of cluster does occur 
and tends to make the cluster more compact, and it may eventually become 
non-fractal at a certain $p$, denoted $p_{SG}$. We can call this transition as 
``sol-gel'' transition~\cite{weitz}; for $p > p_{SG}$, the system will be a 
``gel,'' i.\ e., a fractal cluster, while for $p < p_{SG}$, it will be ``sol,''
characterized by compact and non-fractal clusters. 
If $p_{eff}$ is substantially smaller than 1, there are no longer large
aggregation clusters. Thus we can define the ``melting transition'' 
as the point where, even in the limit of very large $N_m$, 
the aggregation procedure described above leads only to finite clusters 
- there is no ``infinite cluster.''  This transition occurs at a critical
value of $p_{eff}$, denoted $p_M$.
If there is no restructuring effect, this transition is related to the bond
percolation transition, and thus we can consider that $p_M = p_c$, where $p_c$
is the bond percolation threshold (for example, $p_c \sim 0.25$ on a 
$3$-dimensional simple cubic lattice)~\cite{stauffer}.
Next, we discuss the connection between $p_{eff}$ and the physical parameters 
of the real gold/DNA nanoparticle system, as previously analyzed 
in Ref.\ \cite{park1}.

\subsubsection{Determination of $p_{eff}$ by DNA hybridization}

At a low temperature T, a ``link'' is expected to consist 
of a number, say $N_d$, of DNA ``duplexes,'' i.\ e., of pairs
of DNA strands connected to form a molecule.  
In actuality, there is a linker molecule which
emerges from solution to connect two DNA single strands, each
on a different nanoparticle. Even in this case, however,
the melting condition can be reduced to that used here, without
loss of accuracy, as will be shown in Ref.\ \cite{park2}.  Also,
there are actually two chemically distinct DNA single strands
(denoted A and B);
each gold particle has either all A or all B single strands
attached, and the linker molecule can connect only A and B single 
strands.  The existence of two species can be disregarded by 
symmetry when the the concentrations of the two species in 
solution are equal; experiments have been carried out only under 
these conditions~\cite{mirkin,elghanian,storhoff1,others,kiang}.  

Thus, we simply assume that each DNA duplex consists of one 
double strand D, made up of
two short single strands S (each having 12-14 DNA base pairs).
To describe DNA dehybridization, we adopt a simple two-state 
model~\cite{bloomfield,werntges}. 
There also exists a more elaborate
theory which can account for many detailed features of the gold/DNA
nanocomposite, including the dependence of the melting transition 
temperature $T_M$ on the salt concentration of the solvent~\cite{others}. 
However, as mentioned in 
Introduction, both theories reached the same qualitative conclusion 
about the melting transition, namely that the observed sharp 
melting transition and the melting temperature dependence on the system size 
originate from the presence of multiple DNA links between each pair of 
nanoparticles. Moreover, the simple model we use here can be easily extended 
to the similar systems discussed in Ref.~\cite{bruchez,mann,boal,kim}.  

In a simple two-state model, the relative proportion of D and S
is determined by the chemical reaction
\begin{equation}
S + S \rightleftharpoons D.
\label{eq:react}
\end{equation}  
The chemical equilibrium condition corresponding to eq.\ 
(\ref{eq:react}) is
\begin{equation}
\frac{(1 - p(T))^2}{p(T)} = \frac{K(T)}{C_T},
\label{eq:equil}
\end{equation}
where $p(T)$ is the fraction of the
single DNA strands which form double strands by the
reaction (\ref{eq:react}) 
at temperature T, $K(T)$ is a chemical equilibrium constant,
and $C_T$ is the molar concentration of single DNA strands in 
in the sample,
Since $0 < p(T) < 1$,
the physical solution to eq.\ (\ref{eq:equil}) is
\begin{equation}
p(T) = 1 + \frac{1}{2}\left(K^\prime - \sqrt{K^{\prime,2} +
4K^\prime}\right),
\end{equation}
where $K^\prime = K(T)/C_T$.
Since $K^\prime(T)$ is typically an increasing function of T,  
$p(T)$ will generally decrease with increasing T.  In our
calculations, we have also assumed 
the simple van 't Hoff behavior 
\begin{equation}
K(T) = \exp[-\Delta G/k_BT], 
\end{equation}
with a Gibbs free energy of formation
\begin{equation}
\Delta G(T) = c_1(T - T_M^0) + c_3(T - T_M^0)^3,
\end{equation}
choosing the values of $c_1$, $c_3$, and $T_M^0$ to be consistent
with experiments on these DNA molecules\cite{note1}.   The temperature 
$T_M^0$ can be interpreted as the melting temperature of a
single DNA duplex.  

Given $p(T)$, $1 -p_{eff}$ is the probability that {\em none}
of the duplexes forms a double strand.  If the duplexes react
independently, this probability is simply 
\begin{equation}
1 - p_{eff}(T) = [1 - p(T)]^{N_d/z},
\label{eq:peff}
\end{equation}
where $N_d$ is the number of single strands on each monomer, and 
$z$ is the number of nearest neighbors for the given lattice
($z = 6$ for a simple cubic lattice).
Thus, $p_{eff}(T)$ is also a decreasing function of T, but
for $N_d \gg 1$ will typically vary much more sharply with $T$ than
$1 - p(T)$.

The criterion for the melting temperature $T_M$ of the aggregate
is easily written down for a fully occupied lattice.  
If we denote the melting probability by $p_M\equiv p_{eff}(T_M)$, 
the melting temperature for a periodic lattice of monomers is simply
\begin{equation}
1 - p_M = [1 - p(T_M)]^{N_d/z}.
\label{eq:melt}
\end{equation}
Eq.\ (\ref{eq:melt}) implicitly determines $T_M$ in terms of $p_M$,
$N_d$, and $z$.
As discussed in Ref.\cite{park1}, if we assume that the specific links which 
are occupied at temperature $T$ are time-independent,  
the melting takes place at $p_M=p_c$, where $p_c$ is the bond percolation 
threshold for the lattice considered, at which 
an infinite connected path of double DNA strands first forms.
For example, $p_c \sim 0.25$ on a very large simple cubic
lattice.  

Note that, according to eq.\ (\ref{eq:melt}), $p(T_M)$ decreases
with increasing $N_d$ and, hence, with increasing particle radius,
since $N_d$ should be proportional to the surface area of a 
nanoparticle.  To obtain specific values for $T_M$,
we assume $N_d \propto a^2$, set $z = 6$ and use the
experimental result that $N_d = 160$ when $a = 8$ nm~\cite{demers}.
Since $p(T)$ decreases monotonically with $T$, $T_M$ should thus  
{\em increase} monotonically with $a$, as reported in 
experiments\cite{others,kiang}.

In the present paper, we keep our structural model as simple as possible,
so as to focus on the essential features of the
aggregation process.
In reality, there are many complicated issues which should be considered.
For example, the binding of DNA to the gold nanoparticles is a statistical 
process~\cite{zanchet}, and thus there will always be a distribution in 
the number of DNA strands per particle. 
Moreover, because of the high local dielectric function arising from the large
amount of DNA on the particles, not every DNA single strands which is bound 
on the particles can hybridize with the linker DNA~\cite{demers}.
Furthermore, it is unlikely that the gold/DNA composite resulting from the
aggregation process will form a regular crystal.  All these features 
could, however, be included in an extension of our theory.

\section{Optical Properties}

\subsection{Discrete Dipole Approximation}

Given the distribution of clusters, and their geometries, we
calculate their optical properties as follows.  We first assume
that, at any given time $t$, the extinction coefficient $C_{ext}(t)$ of
a given cluster can be calculated as if none of the other
clusters were present.   This amounts to neglecting corrections
due to multiple scattering among the different clusters.
We calculate $C_{ext}$ for each cluster using the so-called
discrete dipole approximation (DDA), first proposed by Purcell
and Pennypacker\cite{pp}.   

As originally formulated, the DDA permits one to calculate
the extinction coefficient of
an irregularly shaped object of complex, frequency-dependent 
dielectric constant $\bar{\epsilon}(\omega)$, embedded in a homogeneous
medium of real dielectric constant $\bar{\epsilon}_h$
and subjected
to an applied electromagnetic wave with electric field
${\bf E}_0\exp({\bf k}\cdot{\bf r} - \omega t) \equiv$
${\bf E}_0({\bf r})e^{-i\omega t}$.  (In our notation, 
the physical field ${\bf E}_{phys}({\bf r}, t)$ is
the real part of this complex quantity.)  In the
DDA, the object is replaced by a collection of $N_{par}$ identical 
point objects with polarizability $\alpha(\omega)$ placed on
a simple cubic lattice having a suitable lattice constant $d$.    

The relation between
the polarizability $\alpha$ of these point objects and
$\epsilon(\omega)$ is discussed in Section IIIB.  
The induced dipole moment ${\bf p}_i$ of the
$i^{th}$ polarizable point object is expressed as
\begin{equation}
\label{eq:2}
{\bf p}_i = \alpha {\bf E}_{loc,i},
\end{equation}
where ${\bf E}_{loc,i}$, the local electric field at the position
of the i$^{th}$ point dipole, is 
\begin{equation}
\label{eq:3}
{\bf E}_{loc,i} = {\bf E}_0\exp[i{\bf k}\cdot{\bf r}_i - i\omega t]
- \sum_{j \neq i}{\bf A}_{ij}\cdot{\bf p}_j.  
\end{equation}
Thus ${\bf E}_{loc,i}$ is the sum of the applied field at
${\bf r}_i$ and all the scattered fields 
$-{\bf A}_{ij}\cdot{\bf P}_j$ emanating from the induced dipoles
at ${\bf r}_j$.  In the DDA, the product 
${\bf A}_{ij}\cdot{\bf p}_j$ can be expressed 
as\cite{lazar,pp,goodman}
\begin{eqnarray}
\label{eq:4}
{\bf A}_{ij}\cdot{\bf p}_j &= & \frac{e^{ikr_{ij}-i\omega t}}{r_{ij}^3}
\{k^2{\bf r}_{ij}\times[({\bf r}_{ij} \times {\bf p}_j]
+ \nonumber \\
& + & \frac{1-ikr_{ij}}{r_{ij}^2}[r_{ij}^2{\bf p}_j -3{\bf r}_{ij}
({\bf r}_{ij}\cdot {\bf p}_j)]\}.
\end{eqnarray}
Here ${\bf r}_{ij} = {\bf r}_i - {\bf r}_j$ and 
$k = \omega/c \equiv 2\pi/\lambda$,
$c$ being the speed of light in vacuum, and $\lambda$ the wavelength
in vacuum.
Eqs.\ (\ref{eq:2})-(\ref{eq:4}) form a coupled set of $3N_{par}$ 
equations, which can be solved for the $N_{par}$ 
dipole moments ${\bf p}_i$
using the complex-conjugate gradient method 
combined with fast Fourier transforms~\cite{goodman}.
Given the ${\bf p}_i$'s, 
$C_{ext}$ for a given cluster is obtained from the relation
\begin{equation}
C_{ext} = \frac{4\pi k}{|{\bf E}_0|^2}Im
\left[\sum_{j = 1}^{N_{par}}{\bf E}_0^*
\exp(-i{\bf k}\cdot{\bf r}_j)\cdot {\bf p}_j\right],
\label{eq:5}
\end{equation}  
where the sum runs over the $N_{par}$ particles in the cluster.

\subsection{Dipole Polarizability}

In the DDA, there are many possible choices for
the relation between the polarizability $\alpha(\omega)$ of the
point objects and $\bar{\epsilon}(\omega)$~\cite{draine}.  
One possible choice is to use 
the Clausius-Mossotti equation:
\begin{equation}
\label{eq:1}
\bar{\epsilon}(\omega) - 1 = \frac{4\pi n\alpha}{1 - (4\pi/3)n\alpha}.
\end{equation}
where $n = 1/d^3$ is the number of point polarizable objects 
per unit volume.  
This relationship between $\bar{\epsilon}(\omega)$ and $\alpha$ is accurate
for point polarizable objects on a cubic mesh, provided that
the wavelength in the medium, $\lambda_m$ is much larger than 
$d$\cite{purcell1}.
However, if $d/\lambda_m$ is not very small, 
this choice can violate the optical theorem.  To prevent this
violation, an extension of the 
Clausius-Mossotti equation to include a radiative reaction correction
has been proposed\cite{CMRC}.  For a simple cubic lattice,
this correction can be incorporated by using a
lattice dispersion relation, which is appropriate for a periodic
cluster~\cite{LDR}. 

One can also choose $\alpha$ by connecting it to
the first scattering coefficient in the Mie theory, usually
denoted $a_1$.  If we take the ``point dipole'' as a sphere of radius
$a$, dielectric constant $\epsilon$, in a host medium of 
dielectric constant $\epsilon_h$, then, as discussed in
Ref.\ \cite{doyle}, the relation
between $\alpha$ and $a_1$ is
\begin{equation}
\alpha=i\frac{3}{2k^3}a_1.\label{eq:a1_term}
\end{equation}
Here 
\begin{equation}
a_1=\frac{m\psi_1(mx)\psi_1'(x)-\psi_1(x)\psi_1'(mx)}
{m\psi_1(mx)\xi_1'(x)-\xi_1(x)\psi_1'(mx)},
\end{equation}
where $\psi_1(x) =  xj_1(x)$ and $\xi_1 = x h_1^{(1)}(x)$,
the complex number 
$m = \sqrt{\epsilon(\omega)/\epsilon_h}$, and 
$x = 2\pi \sqrt{\epsilon}a/\lambda$.
Here $j_1(x)$ is the usual first order spherical Bessel function, and
$h_1^{(1)}$ is the first order spherical Hankel function.

\subsection{Application of the DDA to the Gold/DNA Composite System}

In the present work, each cluster consists of a number of DNA-linked 
individual gold nanoparticles.
The application of DDA to this gold/DNA system clusters has been 
developed extensively for regular clusters~\cite{lazar,lazar1}.
In carrying out the calculations for irregular clusters, we
do not explicitly include the optical 
properties of the DNA molecules, since these absorb primarily in the 
ultraviolet~\cite{storhoff1}.    
We use tabulated values for the complex index of refraction 
$n_{bulk}(\omega)$ of bulk 
gold~\cite{lynch,johnson}, then calculate $C_{ext}$ for each 
cluster using the DDA, 
using a finite-particle-size corrected dielectric function
for the gold particles,'' denoted $\epsilon(\omega)$.

We obtain $\epsilon(\omega)$ by correcting the bulk dielectric
function $\epsilon_{bulk}(\omega) = n_{bulk}^2(\omega)$ to 
account for the additional damping mechanism induced by 
collision of the conduction electrons with the particle surface.
Specifically, we write 
\begin{equation}
\epsilon(\omega) = \epsilon_{bulk}(\omega) + \frac{\omega_p^2}
{\omega(\omega + i/\tau)} - \frac{\omega_p^2}{\omega + i/\tau
+ i/\tau_a},
\label{eq:qse}
\end{equation}
where $\epsilon_{bulk}(\omega)$, $\omega_p$, and $1/\tau$ are
the experimental bulk metal values for the dielectric function,
plasma angular frequency, and relaxation rate.  The quantity
\begin{equation}
1/\tau_a = Av_F/a_{eff}
\label{eq:taua}
\end{equation}
is the surface damping term.
$v_F$ is the Fermi velocity, and $a_{eff}$ is an effective
particle radius, defined by setting $4\pi a_{eff}^3/3 = v_{part}$,
where $v_{part}$ is the particle volume.  (Obviously, this 
expression is exact for spherical particles.)  The constants
$\omega_p$, $\tau$, and $v_F$ are taken from Ref.\ \cite{gh}.
The constant $A$ is a theory-dependent parameter that includes 
details of
the scattering process~\cite{par_A}, and is expected to be of order
unity.
It also can depend on ``chemical interface damping'' 
(i.\ e. transfer of surface plasmon energy into 
excitation modes of the surface metal-adsorbate 
complex)~\cite{persson}.  To obtain the parameter $A$, we
compared our calculations in the dilute limit to  
experimental data for the extinction spectra of 
DNA-linked gold dispersed colloids~\cite{elghanian},
choosing the parameter $A$ as the 
best fitted value, as further discussed below.  

Since we are
treating disordered clusters, we believe that the best
choice for $\alpha$ is eq.\ (\ref{eq:a1_term}), and have used
this equation in our calculations. 
For this choice, the DDA calculation has been
compared to a more accurate
method~\cite{mackowski} for calculating the 
extinction coefficient for a compact spherical 
aggregate of 89 30-nm gold nanospheres in an aqueous medium.
The DDA result was found to be reasonably consistent 
with this more accurate method~\cite{lazar1}.   Also, for a 40-sphere DLA
fractal cluster, Mackowski\cite{mackowski1} has
compared the exact total cross section, as obtained using a
multiple scattering formalism, to that obtained from the
dipole method in combination with eq. (\ref{eq:a1_term}), using
various choices for the sphere index of refraction.
The calculated DDA results agree qualitatively with the
exact multiple-scattering calculation, except for a single
discrepancy, which can be corrected by adding a frequency-independent
constant to the extinction coefficient.

To improve the statistics,
we average the calculated $C_{ext}$ for each cluster over 
possible orientations.  We then sum the averaged extinction 
coefficients of all the individual clusters to obtain
the total extinction coefficient of the suspension.  
This method should be adequate so long as the total volume
fraction of clusters in the suspension is small (dilute regime).

\section{Numerical Results}

We turn now to our numerical results, based on this approach
to modeling the structural and optical properties of gold/DNA
composites.  We begin with the structural properties.

First, we show that our numerical algorithm does indeed
generate an RLCA cluster at $p_{eff} = 1$.  In Fig.\ 1, we
show the radius of gyration $R_g$ for such a cluster, plotted
as a function of $N_m$.  $R_g$ is defined by
the relation
\begin{equation}
R_g^2 = \frac{1}{N_m}\sum_{i=1}^{N_m}|{\bf r}_i - {\bf \bar{r}}|^2,
\label{eq:rg}
\end{equation}
where ${\bf r}_i$ is the position of the i$^{th}$ monomer
and ${\bf \bar{r}} = N_m^{-1}\sum_i{{\bf r}_i}$ is the cluster
center of mass.  $d_f$ is then given by
\begin{equation}
d_f = \lim_{R_g \rightarrow \infty}\frac{d\ln N_m}{d\ln R_g}.
\end{equation}
Fig.\ 1 shows that the log-log plot of $N_m$ against $R_g$ is
indeed a straight line with a slope $d_f = 2.1$, consistent
with expectations for RLCA, provided $N_m$ exceeds about 200.
 
Next, we consider how the radius of gyration $R_g$, and
hence the fractal dimension of the clusters,
varies with $N_{MC}$.  
In Fig.\ 2, we plot $R_g$ as a function of the number of
MC steps $N_{MC}$, for probability $p_{eff}=0.9$.
We consider the evolution of clusters having $N_m$ 
varying from $100$ to $600$.  In several cases, the final result
of the evolution is more than one cluster; in these cases, we
calculate $R_g$ for the largest cluster.  
As shown in Fig.\ 2, as the restructuring of cluster proceeds
with increasing $N_{MC}$, the radius of gyration $R_g$ of the largest
cluster of the system becomes smaller and eventually relaxes to a saturated 
value.  

To characterize these changes, we calculate the fractal dimension 
$d_f$ at the beginning of the simulation, and after the system has fully 
relaxed.  In Fig.\ 3, we show log-log plots of $R_g(N_m)$ 
for two different values of $N_{MC}$, using the data of Fig.\ 2.
For small $N_{MC}$ ($N_{MC} =100$), we measure $d_f \sim 2.1$, 
as expected for an RLCA cluster.  But when we measure the value $d_f$ for the 
saturated $R_g$ at large $N_{MC}$, we find $d_f \sim 3$, corresponding
to compact, non-fractal clusters.  Thus, the clusters are 
becoming more compact with increasing simulation time.  

In Fig.\ 4, we show some typical cluster morphologies corresponding
to the procedure described above.  In Fig.\ 4(a)-(c), we show
a cluster with $N_m = 1000$, at $p = 0.9$, after 0, 7000, and 70000
MC steps. The gradual transition from a fractal morphology to a 
more compact one is evident in the figure.
(This behavior can also be observed at $p=0.95$ although the saturation
time is longer than at $p=0.90$.)  Also during the simulation, 
we observe that most configurations have one large cluster and a few 
monomers as can be seen in Fig. 4 (b) and (c).
This indicates that at large $p_{eff}$
the mechanism for cluster restructuring is mainly
the diffusion of monomers along the surface of the large cluster. 

From these numerical results we may infer some qualitative conclusions
about the sol-gel transition mentioned earlier.  We denote by $\tau_1$
the time needed to break one monomer from the surface of a large cluster.
We estimate this time as $\tau_1 \propto \tau_0/(1 - p_{eff})^{N_{av}}$,
where $\tau_0$ is the time needed to break a single link between
two particles, and $N_{av}$ is the average number of links on one monomer
at the surface of a large cluster.  Thus $\tau_1$ diverges as 
$p_{eff}\rightarrow 1$, but is finite for any $p_{eff} < 1$, allowing
the largest cluster to relax to a non-fractal structure.
Hence, we can expect that $p_{SG}$ should be close to 1, if sufficient
relaxation time is provided. 

Next, we turn to the calculated optical properties of these
gold nanoparticle/DNA composites, based on this structural model.
We begin by showing in Fig.\ 5 the computed extinction coefficients 
as a function of $\lambda$ for gold monomers with 6.5 nm radius, 
including the quantum size corrections embodied in eqs.\ 
(11), (\ref{eq:a1_term}), (\ref{eq:qse}), and (\ref{eq:taua}).  
For comparison, we also show $C_{ext}(\lambda)$ with no quantum 
size correction, and the experimental data 
in Storhoff,\ {\it et\ al.}~\cite{elghanian}.
The rather sharp extinction peak near $\lambda = 530$ nm corresponds
to the well-known surface plasmon peak, in which light is absorbed
by an oscillation of the electronic charge within the gold nanoparticle.  
We find that the quantum-size-corrected extinction
coefficients give the best fit to the experimental data 
if we choose $A=0.85$.  This value is of order unity, as expected.

We now discuss the optical properties of various clusters.  
In Fig.\ 6, we show the specific extinction coefficient 
$C_{ext}(\lambda)/L^3$, as calculated using the DDA for clusters of gold 
monomers of radius $20$ nm, arranged on a simple cubic lattice of edge 
varying from $L = 1$ to $L = 7$ lattice constants, and thus from $N_m = 1$ 
to $N_m = 343$.  We choose the lattice constant $\ell = 48$ nm.
For $L = 1$, the extinction coefficient corresponds to a single
monomer. As $L$ increases, the peak first shifts towards
the infrared, then broadens substantially.  

In Fig.\ 7, we show $C_{ext}(\lambda)/N_m$ as a function of $\lambda$
for $N_m$ varying from $1$ to
$343$, but with particles now forming a single
RLCA cluster.  Once again, the particles have radius 20 nm as
in Fig.\ 6, and the RLCA cluster forms on a simple cubic lattice
of the same lattice constant $\ell = 48$ nm.  The
cluster is generated using the RLCA algorithm described at the
beginning of Sec.\  IIB.  As in Fig.\ 6, the extinction peak is 
red-shifted and broadened
as $N$ increases, but both the shift and the broadening are 
{\em substantially smaller} than in Fig.\ 6.

As a final comparison, we show in Fig.\ 8 the 
specific extinction coefficients 
$C_{ext}(\lambda)/N_m$ for the {\em fully relaxed}
configurations generated by the algorithm of 
Section IIB.  We set $p_{eff} = 0.9$
and consider $N_m$ varying from 1 to 400, as indicated
in the legend.  The behavior of $C_{ext}$ in this case
resembles Fig.\ 6 more than Fig.\ 7; in particular,
the surface plasmon peak is more broadened and
red-shifted than that of Fig.\ 7.  This result is
not surprising, since the clusters of Fig.
8, like those of Fig.\ 6 and unlike those of Fig.\ 7,
are compact and non-fractal, though they incorporate
some disorder.

Figs.\ 6 and 8 show that compact, non-fractal
clusters (whether regular or irregular) have similar
optical properties.  Now, these clusters do have various
structural differences; for example,
the clusters of Fig. 8 have rougher surfaces than
those of Fig. 6.
We conclude that these structural differences do not
significantly affect the cluster extinction coefficients.
Of the three extinction coefficients shown in Figs.\ 6-8, that of Fig.\ 7 
agrees best with experiment; those of Figs.\ 6 and 8 show too large a
broadening and red shift of the surface plasmon peak.  

In Fig.\ 9, we show the calculated normalized extinction 
coefficient $C_{ext}(\lambda)/N_m$ at fixed
wavelength $\lambda = 520$ nm as a function of $p_{eff}$. 
The three continuous curves represent the results of the model
described in our previous paper\cite{park1}; In this model, 
the cluster is constructed starting 
from a simple cubic lattice of linear dimension $L\ell$, fully
occupied by monomers of radius $a$.  Bonds are then
removed with probability $1 - p_{eff}$.  At a given
value of $p_{eff}$, there are one or more clusters, 
depending on the relation of $p_{eff}$ to the bond
percolation threshold $p_c$ for the lattice and on $L$.  
The open circles represent the result of the model 
of this paper - that is, the final cluster is the result of 
a series of MC steps, each consisting of an RLCA 
aggregation, followed by a 
thermally induced breakup of the cluster.  The
calculated $C_{ext}(\lambda)/N_m$ are based on the
long-time and thus relaxed configurations generated 
by this procedure.

As is evident from Fig.\ 9, there is a sharp drop
in $C_{ext}(\lambda)/N_m$ at a characteristic value
of $p_{eff}$ in both models, which we identify with the melting
transition for that model.   For the bond percolation model 
of Ref.\ \cite{park1}, the melting value $p_M$ is
close to 0.25, 
the bond percolation threshold 
on a simple cubic lattice in $d$ = 3.  For the model of the present paper, 
the melting point $p_M$ occurs at a considerably 
higher value of $p_{eff}$. In the present model, unlike the previous
model of Ref.~\cite{park1}, the final cluster results from many steps of 
link removal and reformation. 
Thus the infinite cluster which is formed from the percolation model, can
not survive at the bond percolation threshold because it contains many
weak links which will break apart after several link removal steps.
Thus, there should be an {\em increase} 
in the melting value of $p_M$, as observed in our simulations.

In Fig. 4 (d), we illustrate a typical configuration generated at 
$p_{eff}=0.6$ by the present model.  Here we can clearly see 
that there is no large cluster, even 
though the $p_{eff}$ is much higher than the bond percolation threshold $p_c$.

Even though the value of $p_M$ at melting is larger for 
the present model than in our previous model in Ref.\cite{park1}, 
the {\em sharpness} of the melting transition (as seen in the calculated 
optical properties) is similar to that found previously.  
The reason is that the sharp transition 
results mainly from the individual link properties [cf. eq.
(3)], not the behavior of the cluster.  Thus, the present
model preserves the sharp melting transition reported
in experiments\cite{others,kiang}.
   
In Fig.\ 10, we replot the results of Fig.\ 9,
but with $p_{eff}$ translated into a temperature using
the prescription of eqs.\ (2) and (3).  The higher
melting value of $p_{eff}$ obtained from the present
model now translates into a lower melting temperature $T_M$, 
in comparison to the percolation model in Ref.\ \cite{park1}.

A striking feature of Figs.\ 9 and 10 is the ``rebound effect,'' seen in 
the cross and square of Fig.\ 9, and in the cross in Fig.\ 10.  
The square at $p_{eff} = 1$ in Fig.\ 9 represents the calculated 
$C_{ext}(\lambda)/N_m$ resulting from pure RLCA. The cross 
at $p_{eff} = 0.9$ in Fig.\ 9 or in Fig.\ 10 represents a configuration 
obtained after 7000 MC steps, which is not long enough to
produce the asymptotic compact cluster for this process at
this value of $p_{eff}$.  In both cases, $C_{ext}(\lambda)/N_m$ is larger 
than that of the fully relaxed configuration for this $p_{eff}$.   
This behavior can be understood quite simply.  Since these points
are not fully relaxed, they correspond to clusters which are more fractal 
than the fully relaxed clusters.  Being more fractal, they show some 
characteristics of the melted clusters at this wavelength; hence, they have 
a somewhat higher $C_{ext}(\lambda)/N_m$ than the fully relaxed samples.  
In fact, this rebound effect has been observed in experiments which are
carried out on  gold/DNA 
aggregates~\cite{mirkin,elghanian,storhoff1,others,kiang}.
However, when gold/DNA composites are formed in the presence of a DNA-coated
flat surface, this effect was not observed~\cite{others}.

The observation of the rebound effect in some experiments but
not in others needs to be understood, as does our result that among
the three extinction coefficients shown in Figs.\ 6-8, that of Fig.\ 7,
which represents the fractal aggregate, agrees best with experiment. 
Both results suggests that fractal-like gel state persists up to
surprisingly high temperatures, such as room temperature.  On 
the other hand, our numerical study of our structural model 
suggests that $p_{SG}$ is either 1 or at least close to 1, given
a sufficiently long relaxation time, which implies that a gel state
should exist only a very low temperatures at such long times.
We will discuss both of these questions in the next section.
    
\section{Discussion and Conclusions}

In this paper, we have described a structural model for
the formation of gold/DNA nanocomposites, going
well beyond our earlier, purely percolation model.  The model 
includes several key features expected to play a role
in these aggregates.  For example, we include
both the formation of clusters via reaction-limited cluster-cluster
aggregation, and the thermally induced dehybridization which breaks
up the links between the gold monomers.   We also calculate
the aggregate extinction coefficient $C_{ext}(\lambda)$
for some of the model structures.

Our structural model is characterized by two types of transitions.
The first is the melting transition mentioned above.   Above $p_M$,
in the limit of a large number of monomers, the system is 
characterized by at least one large cluster; below $p_M$, this
cluster breaks up into two or more smaller clusters.  In our
previous percolation model, $p_M$ coincides with the percolation
threshold, but in the present model, it occurs at a higher value
of $p_{eff}$.  It is observable as a rather sharp transition in
$C_{ext}(\lambda)$ at a characteristic wavelength of about $520$ nm.  
The other transition is the sol-gel transition.  
For $p > p_{SG}$, the aggregate has a fractal gel-like structure,
while for $p < p_{SG}$, it is a non-fractal sol.
We have presented some numerical evidence showing that this 
sol-gel transition will occur at $p_{eff}\sim 1$,
 if the sufficient relaxation time is provided.

However, our calculated $C_{ext}(\lambda)$
agree best with experiment if we assume that the composites grow
into a fractal gel-like structure, rather than the compact sol
structure.   Specifically, the surface plasmon peak seen in
$C_{ext}(\lambda)$ for isolated gold monomers is broadened and
red-shifted by about the right amount, in comparison to experiment,
for the fractal clusters, but by far too much for the compact,
non-fractal clusters.   

Thus, it appears that the full restructuring of our model, which would
eventually give rise to a non-fractal, does not go to completion
in the experimental composites.  There are various possible reasons
for this behavior.   Most likely, the time required for a full
relaxation is simply too long.  If the composite is well below
its melting temperature, only a small fraction of links are broken
and the relaxation is likely to be very slow.  As an illustration,
we found that, for $p_{eff} = 0.9$, even $N_{MC} = 7000$ is not
long enough to produce the relaxation necessary to generate a
non-fractal cluster.  

Another suggestive piece of experimental evidence is the rebound effect.  
According to our calculations, $C_{ext}(\lambda)$ 
should show a characteristic increase at $\lambda = 520$ nm, 
above its fully relaxed value,
when the cluster is relaxed for only $N_{MC} = 7000$ steps.
With this increase, $C_{ext}(\lambda)$ is slightly larger than
the fully relaxed value expected at this $\lambda$ for a non-fractal
cluster.  Experiments do indeed show indications of this rebound 
effect~\cite{mirkin,elghanian,storhoff1,others,kiang}.

In summary, we have presented a model
for the structural evolution of gold/DNA composites.  This model leads to
a wide range of structures which depend on the characteristic
parameters of the model: the number of monomers, the simulation time,
and the temperature.  The optical properties of these composites
calculated from the resulting structure are consistent with 
experiment.  In particular, we obtain the observed 
sharp melting transition with a characteristic dependence on 
monomer radius, a characteristic
shift and broadening of the surface plasmon peak in the extinction,
of about the experimental magnitude, and a characteristic rebound in
$C_{ext}(\lambda)$, which indicates that the restructuring of
the experimental composites has not been able to go to completion.
In view of these successes, we plan to extend the present work to
study the sol-gel transition in gold/DNA nanocomposites.
The results of this investigation will be described in a future
publication\cite{park3}.

\section{Acknowledgments}

This work has been supported by grant NSF DMR01-04987, and by
an Ohio State University Postdoctoral Fellowship. 
Calculations were carried out using the facilities of the Ohio Supercomputer
Center.
We thank Prof. M. Kardar, Prof. C. H. Kiang, Prof. D. R. Nelson, and 
Prof. D. A. Weitz for valuable discussions; SYP also thanks
Prof. M. Kardar for his kind hospitality during the stay in MIT.  

\begin{figure}
\epsfxsize=8cm \epsfysize=6cm \epsfbox{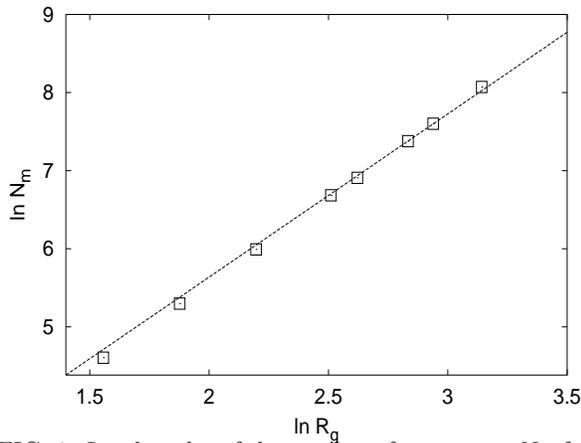}
\caption{ Log-log plot of the number of monomers $N_m$ for an
RLCA cluster at $p_{eff} = 1$, plotted as a function of the radius of 
gyration $R_g$.  Squares: present calculations; dashed line:
least squares fit.  The slope of the dashed line 
is $2.1 = d_f$, consistent with expectations for RLCA clusters in 
$d = 3$.
}
\end{figure}

\begin{figure}
\epsfxsize=8cm \epsfysize=6cm \epsfbox{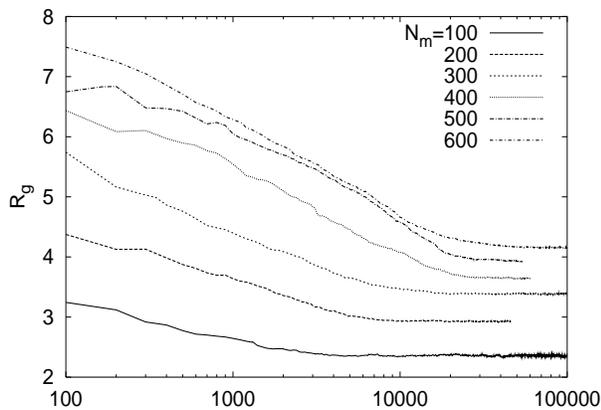}
\caption{  Variation of $R_g$ with number of
Monte Carlo steps $N_{MC}$, as obtained using the algorithm 
described in the text for probability $p_{eff}=0.9$,
and clusters with a number of monomers $N_m$ varying from
100 to 600, as indicated.
}
\end{figure}

\begin{figure}
\epsfxsize=8cm \epsfysize=6cm \epsfbox{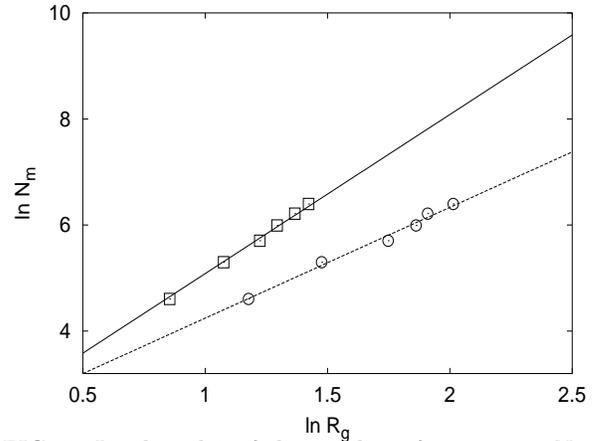}

\caption{ Log-log plot of the number of monomers $N_m$ for the
clusters generated by the algorithm described in the text, 
at $p_{eff} = 0.9$, as a function of the radius of gyration
$R_g$. 
The open circles represent the data of Fig.\ 2 for $N_m$ as a function
of $R_g$ at $N_{MC}=100$.  The slope of dashed line is 
$2.1$, the expected value for RLCA.  The open squares represent the 
relaxed values of $R_g$ for large $N_{MC}$, also taken 
from Fig.\ 2.  They are well fitted by the
solid line, which has a slope of $3.0$,
corresponding to a compact, non-fractal cluster.
}
\end{figure}

\begin{figure}
\epsfxsize=8cm \epsfysize=6cm \epsfbox{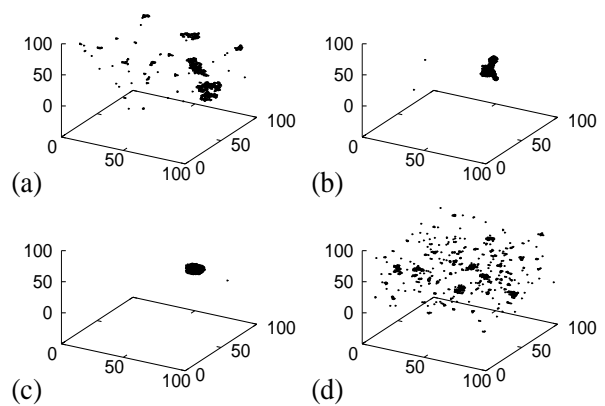}
 
\caption{ (a) Initial form of the cluster, after the first reaction-limited
aggregation ($N_{MC} = 0$), for $N_m = 1000$, $p = 0.9$.  The
clusters are divided from an RLCA fractal, with fractal dimension 
$d_f \sim 2.1$.
(b). Same cluster, but after 7000 Monte Carlo steps.  (c) Same
as (b), but after 70000 Monte Carlo steps (nearly saturated).
(d). Clusters with $p = 0.6$, $N_m = 1000$, and $N_{MC} = 6000$.
}
\end{figure}

\begin{figure}
\epsfxsize=8cm \epsfysize=6cm \epsfbox{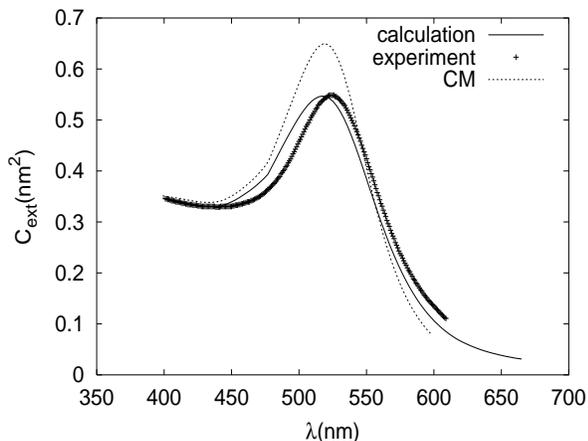}
\caption{Extinction coefficient $C_{ext}(\lambda)$ per unit
volume of gold for a dilute suspension of gold nanoparticles in water, 
plotted for a particle radius $a=6.5$ nm.  Crosses: experimental
data of Ref.\ [9].  
Full curve: calculated extinction coefficient including quantum
size corrections, as computed from eqs.\ (\ref{eq:qse}) 
and (\ref{eq:taua}) with $A = 0.85$.  Dashed line: calculated extinction
coefficient without quantum size corrections.}
\end{figure}

\begin{figure}
\epsfxsize=8cm \epsfysize=6cm \epsfbox{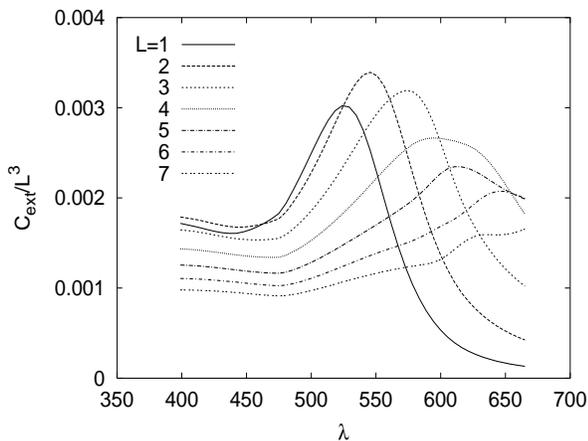}
\caption{$C_{ext}(\lambda)/L^3$ plotted as a function of $\lambda$
for $L \times L \times L$ clusters with edge varying from
$L = 1$ to $L = 7$ lattice constants.
In all cases, we assume the cluster forms a cubic compact cluster 
with particles of radius 20 nm, lattice constant $\ell = 48$ nm; 
so the number $N_m$ of monomers
varies from $1$ to $343$.  We use the DDA including quantum
size corrections from eqs.\ (\ref{eq:qse}) and (\ref{eq:taua})
and polarizability obtained from eqs.\ (10), (11), (13), and (14)}
\end{figure}

\begin{figure}
\epsfxsize=8cm \epsfysize=6cm \epsfbox{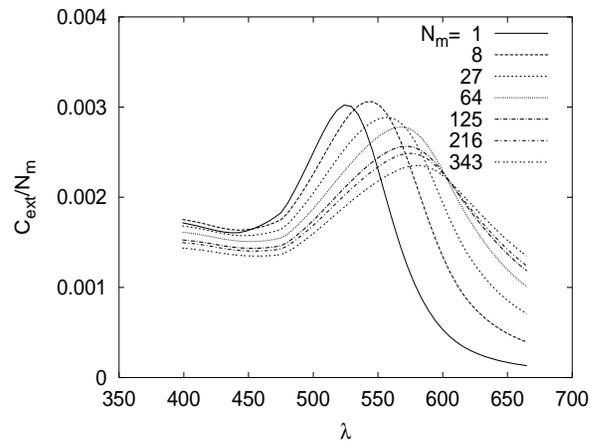}

\caption{ $C_{ext}(\lambda)/N_m$ plotted as a function of $\lambda$
for various number of particles $N_m$ from $1$ to $343$.
In all cases, we assume the cluster forms a RLCA cluster with
particles of radius 20 nm, and is generated on a lattice with
lattice constant $\ell = 48$ nm.  $C_{ext}(\lambda)$ is calculated
using the DDA as in Fig.\ 5.
}
\end{figure}

\begin{figure}
\epsfxsize=8cm \epsfysize=6cm \epsfbox{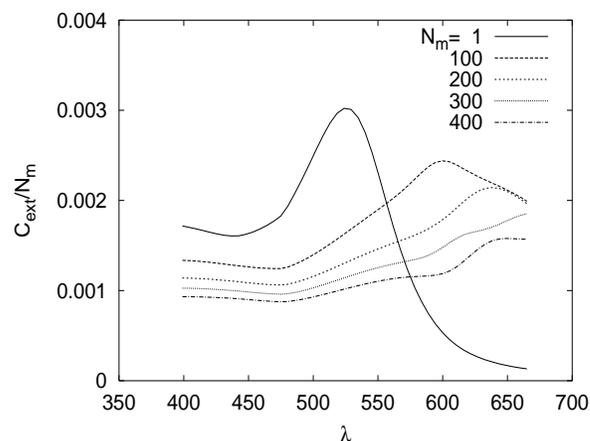}

\caption{ $C_{ext}(\lambda)/N_m$ plotted as a function of $\lambda$
for a number $N_m$ of monomers varying from $1$ to $400$.
In all cases, we use relaxed configurations 
from the algorithm in the text, whose radii of gyrations are
shown in Fig.\ 2.  (Examples of these configurations are
shown in Fig.\ 4.)
Once again, we assume a monomer radius is 20 nm, a lattice constant
of 48 nm, and
use the same algorithm as in Figs.\ 5 and 6 to calculate
the extinction coefficient.  
}
\end{figure}

\begin{figure}
\epsfxsize=8cm \epsfysize=6cm \epsfbox{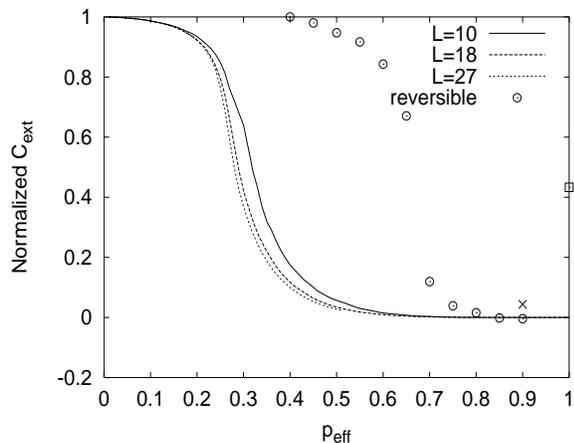}

\caption{ Normalized $C_{ext}(\lambda)$ at a fixed wavelength of
$\lambda = 520$ nm, as calculated in the DDA, for clusters
of various geometries and different numbers $N_m$ of monomers
of radius 20 nm, plotted as a function of $p_{eff}$. 
Full curves: clusters generated by percolation model 
of Ref.\ [18]; 
$p_{eff}$ represents the fraction of links which
are present in this model.  The total number of monomers is
$L^3$.  Open circles: $C_{ext}$ at $\lambda = 520$ nm for the model of the
present paper, with $N_m$ = 1000, with fully relaxed long-time configurations.
The square at $p_{eff}=1.0$ represents the calculated
$C_{ext}$ at $\lambda = 520$ nm assuming RLCA clusters. 
Cross at $p_{eff}=0.9$ represent the calculated values of $C_{ext}$ 
at $\lambda = 520$ nm for the unsaturated configuration obtained after 
$N_{MC}=7000$ steps; the cross at $p_{eff} = 1$ is the value of $C_{ext}$
obtained after a single RLCA aggregation.}
\end{figure}

\begin{figure}
\epsfxsize=8cm \epsfysize=6cm \epsfbox{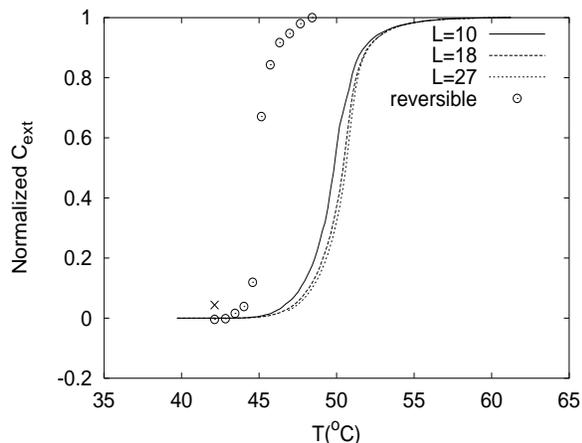}

\caption{ Same as Fig.\ 8, but with $p_{eff}$ translated into
a temperature, using eqs.\ (2) and (3).  The cross
corresponds to $p_{eff} = 0.9$ after $N_{MC} = 7000$.
}
\end{figure}

\end{multicols}

\begin{references}
\bibitem{note} In reality, $\tau_{bind}$ will not remain unchanged
as the aggregation proceeds, because
the number of aggregation attempts depends on the number of 
clusters at a given stage of the simulation.  This number
become smaller as aggregation proceeds.  We do not consider this effect
in our simulations.
\bibitem{note1} In principle, there should also be a term
of the form $c_2(T - T_M^0)^2$ in this expansion.  However,
experiments on isolated pairs of DNA molecules appear to be
best fit by $c_2 = 0$.  We have therefore continued to use
$c_2 = 0$ in our treatment of the aggregate melting.
\end{references}
\end{document}